# Improved three-dimensional color-gradient lattice Boltzmann model for immiscible multiphase flows

Z. X. Wen, Q. Li*, and Y. Yu

*School of Energy Science and Engineering, Central South University, Changsha 410083, China*

Kai. H. Luo

*Department of Mechanical Engineering, University College London, Torrington Place, London WC1E 7JE, UK*

*Corresponding author: qingli@csu.edu.cn

**Abstract**

In this paper, an improved three-dimensional color-gradient lattice Boltzmann (LB) model is proposed for simulating immiscible multiphase flows. Compared with the previous three-dimensional color-gradient LB models, which suffer from the lack of Galilean invariance and considerable numerical errors in many cases owing to the error terms in the recovered macroscopic equations, the present model eliminates the error terms and therefore improves the numerical accuracy and enhances the Galilean invariance. To validate the proposed model, numerical simulation are performed. First, the test of a moving droplet in a uniform flow field is employed to verify the Galilean invariance of the improved model. Subsequently, numerical simulations are carried out for the layered two-phase flow and three-dimensional Rayleigh-Taylor instability. It is shown that, using the improved model, the numerical accuracy can be significantly improved in comparison with the color-gradient LB model without the improvements. Finally, the capability of the improved color-gradient LB model for simulating dynamic multiphase flows at a relatively large density ratio is demonstrated via the simulation of droplet impact on a solid surface.

# I. Introduction

In the past three decades, the lattice Boltzmann (LB) method [1-9], which originates from the lattice gas automaton (LGA) method [10], has been developed into an efficient numerical approach for simulating fluid flow and heat transfer. Different from conventional numerical methods, which are based on the direct discretization of macroscopic governing equations, the LB method is built on the mesoscopic kinetic equation. It tracks the evolution of a particle distribution function and then accumulates the particle distribution function to obtain the macroscopic properties. Owing to its kinetic nature, the LB method exhibits some advantages over conventional numerical methods. For example, in the LB equation the convective operator (the streaming process) is linear, whereas the convective terms of the Navier-Stokes equations are nonlinear [11]. Moreover, in the LB simulations the complex boundary conditions can be formulated with the elementary mechanical rules such as the bounce-back rule according to the interaction of the "LB particles" with the solid walls. Furthermore, the LB method is ideal for parallel computing because of its explicit scheme and the local interactions.

Since the emergence of the LB method, its applications to multiphase flows have always been a very important theme of this method and various multiphase LB models have been developed from different points of view [12]. Generally, most of the existing multiphase LB models can be classified into the following four categories [5-7], i.e., the color-gradient LB method, the pseudopotential LB method, the free-energy LB method, and the phase-field LB method. The first color-gradient LB model was proposed by Gunstensen *et al*. [13], which is also the earliest mulitcomponent extension of the LGA method to the LB method [14]. In the color-gradient LB method, two distribution functions are introduced to represent two different fluids and a color-gradient-based perturbation operator is employed to generate the surface tension as well as a recoloring step for separating different phases or components. The pseudopotential LB method, which is the simplest multiphase LB method, was introduced by Shan and Chen [15,16]. In this method, the fluid interactions are mimicked by an interparticle potential, through which the separation of different phases or components can be achieved naturally. The free-energy LB method was developed by Swift *et al*. [17,18] based on thermodynamics considerations. They proposed to modify the second-order moment of the particle equilibrium distribution function so as to include a non-ideal thermodynamic pressure tensor. The phase-field LB method is based on the phase-field theory, in which the interface dynamics is described by an order parameter that obeys the Cahn-Hilliard equation or a Cahn-Hilliard-like equation [19].

Each of these multiphase LB methods has its advantages and limitations. A comprehensive review of the pseudopotential LB method and the phase-field LB method can be found in Ref. [7]. In addition, the book by Huang, Sukop and Lu [12] is also dedicated to the multiphase LB methods. In this work, we restrict our study to the color-gradient multiphase LB method, which exhibits very low dissolution for tiny droplets or bubbles [20] in comparison with other multiphase LB methods. In the original color-gradient LB model devised by Gunstensen *et al*. [13], the work done by the color gradient against the color flux was maximized to force the colored particles to move towards fluids with the same color. In addition, the model of Gunstensen *et al.* suffers from the limitation of equal densities for two-phase flows. Some improvements have been conducted to overcome the shortcomings of the original color-gradient model. Grunau *et al*. [21] modified the form of the particle equilibrium distribution function to allow for variable density and viscosity ratios. Latva-Kokko and Rothman [22] replaced the numerical maximization recoloring step of Gunstensen *et al.*'s model with a formulaic segregation algorithm, which solves the lattice pinning problem at the interface region and significantly improves the computational efficiency of the color-gradient LB method.

Later, Reis and Phillips [23] proposed a new perturbation operator for generating the surface tension of the color-gradient LB method and derived a theoretical expression for the surface tension through its mechanical definition. Liu *et al*. [24] extended the model of Reis and Philips to three-dimensional space by deriving a generalized perturbation operator, in which an expression for the surface tension parameter is directly obtained without approximations. However, similar to the free-energy multiphase LB method, the color-gradient multiphase LB method also modifies the equilibrium distribution function [23,24] to incorporate the pressure of fluid. Hence it also suffers from the lack of Galilean invariance [7]. Through the Chapman-Enskog analysis, Huang *et al.* [25] showed that some error terms exist in the macroscopic momentum equation recovered from the color-gradient multiphase LB method. They demonstrated that for two-phase flows with different densities the error terms significantly affect the numerical accuracy. A scheme has been proposed by Huang *et al.* [25] to eliminate the error terms, but they emphasized that their scheme just works well for cases of density ratios less than 10.

Recently, Ba *et al.* [26] developed a two-dimensional multiple-relaxation-time (MRT) color-gradient LB model for multiphase flows. To eliminate the error terms in the macroscopic momentum equation, an extension of Li *et al.*'s approach [27] was made, which was devised for recovering $p = \rho RT$ in a double-distribution-function LB model on standard lattices for thermal compressible flows. In the present work, we aim at proposing an improved three-dimensional color-gradient LB model for multiphase flows. The error terms in the momentum equation are removed following the approach of Li *et al.* [27]. To be specific, a high-order term is added to the equilibrium distribution function, through which the off-diagonal elements of the third-order moment of the equilibrium distribution function satisfy the required relationship for recovering the Navier-Stokes equations. Meanwhile, the deviations of the diagonal elements are corrected through introducing a correction term into the LB equation. The rest of the present paper is organized as follows. In Sec. II, the existing three-dimensional color-gradient LB models are briefly introduced. The improved three-dimensional color-gradient LB model is proposed in Sec. III. Numerical simulations are performed in Sec. IV to validate the improved model. Finally, Sec. V concludes the present paper.

## II. The existing 3D color-gradient LB models

### A. The color-gradient LB equation

In the color-gradient LB method, the two immiscible fluids are represented by a red fluid and a blue fluid, respectively. The corresponding distribution functions are denoted by $f_i^k$, where $i$ is the lattice velocity direction and $k = R$ or $B$ denotes the color ("Red" or "Blue"). The total distribution function is defined as $f_i = f_i^R + f_i^B$. The evolution of the distribution functions is governed by the following LB equation [28]:

$$f_i^k\left(\mathbf{x}+\mathbf{e}_i\delta_t, t+\delta_t\right) - f_i^k\left(\mathbf{x}, t\right) = \Omega_i^k\left(\mathbf{x}, t\right), \tag{1}$$

where $\mathbf{x}$ is the spatial position, $\mathbf{e}_i$ is the discrete velocity in the $i$th direction, $t$ is the time, $\delta_t$ is the time step, and $\Omega_i^k$ is the collision operator [23,28]

$$\Omega_i^k = (\Omega_i^k)^{(3)}\left[(\Omega_i^k)^{(1)} + (\Omega_i^k)^{(2)}\right], \tag{2}$$

where $(\Omega_i^k)^{(1)}$ is the single-phase collision operator, $(\Omega_i^k)^{(2)}$ is the perturbation operator, which is used to generate the surface tension, and $(\Omega_i^k)^{(3)}$ is the recoloring operator responsible for phase segregation and maintaining the phase interface [24,26]. When the Bhatnagar-Gross-Krook (BGK) collision operator is applied, the single-phase collision operator is given by

$$(\Omega_i^k)^{(1)} = -\frac{1}{\tau}\left(f_i^k\left(\mathbf{x}, t\right) - f_i^{k,eq}\left(\mathbf{x}, t\right)\right), \tag{3}$$

where $\tau$ is the non-dimensional relaxation time and $f_i^{k,eq}$ is the equilibrium distribution function of $f_i^k$. The macroscopic variables are calculated by

$$\rho_R = \sum_i f_i^R, \quad \rho_B = \sum_i f_i^B, \quad \rho = \sum_k \rho_k, \quad \rho\mathbf{u} = \sum_i\sum_k \mathbf{e}_i f_i^k, \tag{4}$$

where $\rho_k$ is the density of fluid $k$, $\rho$ is the total density, and $\mathbf{u}$ is the macroscopic velocity.

### B. 3D color-gradient LB models

The first three-dimensional color-gradient LB model is attributed to Tölke *et al.* [28], who constructed a three-dimensional nineteen-velocity (D3Q19) color-gradient LB model for immiscible multiphase flows based on the studies of Gunstensen *et al.* [13] and Grunau *et al.* [21]. The lattice velocities $\{\mathbf{e}_i\}$ of the D3Q19 lattice are given by

$$\mathbf{e}_i = c\begin{bmatrix} 0 & 1 & -1 & 0 & 0 & 0 & 0 & 1 & -1 & 1 & -1 & 1 & -1 & 1 & -1 & 0 & 0 & 0 & 0 \\ 0 & 0 & 0 & 1 & -1 & 0 & 0 & 1 & -1 & -1 & 1 & 0 & 0 & 0 & 0 & 1 & -1 & 1 & -1 \\ 0 & 0 & 0 & 0 & 0 & 1 & -1 & 0 & 0 & 0 & 0 & 1 & -1 & -1 & 1 & 1 & -1 & -1 & 1 \end{bmatrix}, \tag{5}$$

where $c = 1$ is the lattice constant. The equilibrium distribution function is chosen as [28]

$$f_i^{k,eq} = \rho_k\left(\phi_i^k + \omega_i\left[\frac{3}{c^2}(\mathbf{e}_i\cdot\mathbf{u}) + \frac{9}{2c^4}(\mathbf{e}_i\cdot\mathbf{u})^2 - \frac{3}{2c^2}|\mathbf{u}|^2\right]\right), \tag{6}$$

where $\omega_i$ is given by $\omega_0 = 1/3$, $\omega_{1-6} = 1/18$, and $\omega_{7-18} = 1/36$, and $\phi_i^k$ is employed to incorporate the pressure of fluid $k$, i.e., $p_k = \rho_k (c_s^k)^2$. The following perturbation operator $(\Omega_i^k)^{(2)}$ is adopted [28]:

$$(\Omega_i^k)^{(2)} = A|\mathbf{C}|\left(\frac{(\mathbf{e}_i\cdot\mathbf{C})^2}{|\mathbf{C}|^2} - \frac{5}{9}\right), \tag{7}$$

where $\mathbf{C}$ is the color gradient and the free parameter $A$ is proportional to the surface tension.

Another three-dimensional color-gradient LB model can be found in the study of Liu *et al.* [24], who extended the perturbation operator proposed by Reis and Phillips [23] to three-dimensional space

$$(\Omega_i^k)^{(2)} = \frac{A_k}{2}|\nabla\rho^N|\left[\omega_i\frac{(\mathbf{e}_i\cdot\nabla\rho^N)^2}{|\nabla\rho^N|^2} - B_i\right], \tag{8}$$

where $\rho^N = \left(\rho_R/\rho_R^{\text{in}} - \rho_B/\rho_B^{\text{in}}\right)/\left(\rho_R/\rho_R^{\text{in}} + \rho_B/\rho_B^{\text{in}}\right)$ with $\rho_R^{\text{in}}$ and $\rho_B^{\text{in}}$ being the initial densities of the red and blue fluids, respectively, and $B_i$ in Eq. (8) is given by $B_0 = -1/3$, $B_{1-6} = 1/18$, and $B_{7-18} = 1/36$. The above perturbation operator leads to the following surface tension [24]:

$$\sigma = \frac{2}{9}(A_R + A_B)\tau c^4\delta_t. \tag{9}$$

Moreover, Liu *et al.* [24] employed the recoloring algorithm proposed by Latva-Kokko and Rothman [22], which can solve the lattice pinning problem and reduce the spurious velocities. According to the recoloring algorithm of Latva-Kokko and Rothman, the recoloring steps for the red and blue fluids can be defined as follows [22,29]:

$$f_i^{R,+} = \frac{\rho_R}{\rho}f_i^* + \beta\frac{\rho_R\rho_B}{\rho^2}\cos(\varphi_i)\sum_k f_i^{k,eq}(\rho_k, \alpha_k, \mathbf{u}=0), \tag{10}$$

$$f_i^{B,+} = \frac{\rho_B}{\rho} f_i^* - \beta \frac{\rho_R \rho_B}{\rho^2} \cos(\varphi_i) \sum_k f_i^{k,eq}(\rho_k, \alpha_k, \mathbf{u} = 0), \tag{11}$$

where $\beta$ is a free parameter controlling the interface thickness, $\varphi_i$ is the angle between the color gradient $\nabla \rho^N$ and the lattice direction $\mathbf{e}_i$, which yields $\cos(\varphi_i) = (\mathbf{e}_i \cdot \nabla \rho^N) / (|\mathbf{e}_i| |\nabla \rho^N|)$, and $f_i^*$ is the post-perturbation value of the total distribution function, namely $f_i^* = \sum_k f_i^{*,k}$, in which $f_i^{*,k}$ is

$$f_i^{*,k}(\mathbf{x}, t) = f_i^k(\mathbf{x}, t) + (\Omega_i^k)^{(1)}(\mathbf{x}, t) + (\Omega_i^k)^{(2)}(\mathbf{x}, t). \tag{12}$$

With Eqs. (10) and (11), the "streaming" process is implemented as $f_i^k(\mathbf{x} + \mathbf{e}_i \delta_t, t + \delta_t) = f_i^{k,+}(\mathbf{x}, t)$. In the study of Liu *et al.* [24], the equilibrium distribution function is also defined by Eq. (6) but $\phi_i^k$ is

$$\phi_i^k = \begin{cases} \alpha_k, & i = 0, \\ (1 - \alpha_k)/12, & i = 1, \cdots, 6, \\ (1 - \alpha_k)/24, & i = 7, \cdots, 18, \end{cases} \tag{13}$$

which corresponds to the pressure $p_k = \rho_k (c_s^k)^2 = 0.5 \rho_k c^2 (1 - \alpha_k)$.

### C. The error terms

The error terms in the momentum equation recovered from the three-dimensional color-gradient LB models have been identified by Huang *et al.* [25] through the Chapman-Enskog analysis and are given by

$$U_\alpha^k = \partial_\beta \left\{ \left( \tau - \frac{1}{2} \right) \left( \frac{1}{3} c^2 - (c_s^k)^2 \right) \left[ u_\beta \partial_\alpha \rho_k + u_\alpha \partial_\beta \rho_k + \partial_\gamma (\rho_k u_\gamma) \delta_{\alpha\beta} \right] \right\}, \tag{14}$$

where the subscripts $\alpha$, $\beta$, and $\gamma$ denote the $x$, $y$, or $z$ coordinate and $\delta_{\alpha\beta}$ is the Kronecker delta. For two-phase flows with identical densities, $\phi_i^k = \omega_i$ is usually adopted and then $f_i^{k,eq}$ given by Eq. (6) reduces to the standard equilibrium distribution function in the LB method, which leads to $(c_s^k)^2 = c^2/3$. Accordingly, the error terms disappear for two-phase flows with identical densities. However, for two-phase flows with different densities, the error terms in Eq. (14) will make the Galilean invariance lost and may affect the numerical accuracy significantly since the density gradient cannot be neglected near the interface.

Recently, Saito *et al.* [30] constructed a three-dimensional 27-velocity (D3Q27) color-gradient LB

model, in which an enhanced equilibrium distribution function devised by Leclaire *et al.* [31] is adopted

$$f_i^{k,eq} = \rho_k \left( \phi_i^k + \omega_i \left[ \frac{3}{c^2}(\mathbf{e}_i \cdot \mathbf{u}) + \frac{9}{2c^4}(\mathbf{e}_i \cdot \mathbf{u})^2 - \frac{3}{2c^2}|\mathbf{u}|^2 \right] \right) + \Phi_i^k , \tag{15}$$

where $\Phi_i^k$ is an additional term, which was originally employed by Che Sidik and Takahiko [32] for a free-energy LB model and was extended to the color-gradient LB method by Leclaire *et al.* [31]. Nevertheless, it is noticed that both Che Sidik and Takahiko [32] and Leclaire *et al.* [31] showed that there are still some error terms in the recovered macroscopic momentum equation, which can be found in Eqs. (29)-(33) of Ref. [31]. The main error terms are similar to the aforementioned error terms given by Eq. (14).

## Ⅲ. Improved 3D color-gradient LB model

### A. Theoretical analysis

In this section, the physical origin of the error terms in Eq. (14) is analyzed. Taking the second-order and third-order moments of the equilibrium distribution function given by Eq. (6), we can find that

$$\sum_i e_{i\alpha} e_{i\beta} f_i^{k,eq} = \rho_k u_\alpha u_\beta + \rho_k (c_s^k)^2 , \tag{16}$$

$$\sum_i e_{i\alpha} e_{i\beta} e_{i\gamma} f_i^{k,eq} = \rho_k \frac{c^2}{3}\left( u_\alpha \delta_{\beta\gamma} + u_\beta \delta_{\alpha\gamma} + u_\gamma \delta_{\alpha\beta} \right). \tag{17}$$

As seen in Eq. (16), the usual pressure $p = \rho_k c^2/3$ has been replaced by $p_k = \rho_k (c_s^k)^2$. However, in the third-order moment given by Eq. (17), the pressure is still defined as $p = \rho_k c^2/3$. Through the Chapman-Enskog analysis, it can be found that the error terms in Eq. (14) just arise from such an inconsistency. If $\rho_k c^2/3$ in Eq. (17) can be replaced by $\rho_k (c_s^k)^2$, the error terms can be removed. However, the symmetry of the standard lattices (such as the D2Q9, D3Q19, and D3Q27 lattices) is insufficient to completely support the replacement of $\rho_k c^2/3$ in Eq. (17) with $\rho_k (c_s^k)^2$.

Fortunately, the off-diagonal elements of the third-order moment of the equilibrium distribution function can satisfy the required relationship by adding a high-order term to the equilibrium distribution function, as shown in Ref. [27] for recovering $p = \rho RT$ in a compressible LB model on standard lattices. Following the study of Li *et al.* [27], the new equilibrium distribution function is defined as

$$f_i^{k,eq} = \rho_k \left( \phi_i^k + \omega_i \left[ \frac{3}{c^2}(\mathbf{e}_i \cdot \mathbf{u}) + \frac{9}{2c^4}(\mathbf{e}_i \cdot \mathbf{u})^2 - \frac{3}{2c^2}|\mathbf{u}|^2 + \frac{3(\mathbf{e}_i \cdot \mathbf{u})}{2c^2}\left( \frac{3(c_s^k)^2}{c^2} - 1 \right)\left( \frac{3|\mathbf{e}_i|^2}{c^2} - 5 \right) \right] \right), \quad (18)$$

which yields

$$\sum_i e_{i\alpha} e_{i\beta} e_{i\gamma} f_i^{k,eq} = \begin{cases} \rho_k \left(c^2/3\right)\left(u_\alpha \delta_{\beta\gamma} + u_\beta \delta_{\alpha\gamma} + u_\gamma \delta_{\alpha\beta}\right), & \text{if } \alpha = \beta = \gamma, \\ \rho_k (c_s^k)^2 \left(u_\alpha \delta_{\beta\gamma} + u_\beta \delta_{\alpha\gamma} + u_\gamma \delta_{\alpha\beta}\right), & \text{others}. \end{cases} \quad (19)$$

For the off-diagonal elements of the third-order moment of $f_i^{k,eq}$, it can be seen that $\rho_k c^2/3$ has been replaced by $\rho_k (c_s^k)^2$. Nevertheless, the diagonal elements ($\alpha = \beta = \gamma$) still deviate from the required relationship owing to the low symmetry of the standard lattices.

To remove the error terms caused by the diagonal elements of the third-order moment of $f_i^{k,eq}$, a correction term can be added to the single-phase collision operator

$$(\Omega_i^k)^{(1)} = -\frac{1}{\tau}\left(f_i^k(\mathbf{x},t) - f_i^{k,eq}(\mathbf{x},t)\right) + \delta_t \left(1 - \frac{1}{2\tau}\right) G_i^k(\mathbf{x},t), \quad (20)$$

where $G_i^k$ is the correction term and the coefficient $(1 - 0.5/\tau)$ in front of $G_i^k$ is responsible for eliminating the discrete effect of a forcing or source term in the LB equation [33]. The zeroth- and first-order moments of the correction term satisfy the following relationships:

$$\sum_i G_i^k = 0, \quad \sum_i \mathbf{e}_i G_i^k = 0. \quad (21)$$

The constraints on the second-order moment of the correction term can be derived through the Chapman-Enskog analysis, which can be implemented by introducing the following multi-scale expansions [34]:

$$f_i^k = f_i^{k,(0)} + \varepsilon f_i^{k,(1)} + \varepsilon^2 f_i^{k,(2)} + \cdots, \quad (22)$$

$$\partial_t = \varepsilon \partial_{t_1} + \varepsilon^2 \partial_{t_2}, \quad \nabla = \varepsilon \nabla_1, \quad G_i^k = \varepsilon G_{1i}^k, \quad (23)$$

where $\varepsilon$ is the expansion parameter.

According to the studies of Reis and Phillips [23] and Liu *et al.* [24], the recoloring step is not considered in the Chapman-Enskog analysis, and therefore the recoloring operator $(\Omega_i^k)^{(3)}$ can be regarded as a unit operator. Meanwhile, the perturbation operator $(\Omega_i^k)^{(2)}$ only affects the surface tension term and has been well demonstrated in Refs. [23,24]. Therefore, in the present study, the Chapman-Enskog analysis is performed for Eq. (1) with $\Omega_i^k(\mathbf{x},t)$ being treated as $\Omega_i^k = (\Omega_i^k)^{(1)}$.

Taking the Taylor-series expansion of the left-hand side of Eq. (1) and using the multi-scale expansions given by Eqs. (22) and (23), we can rewrite the LB equation in the consecutive orders of $\varepsilon$ as follows:

$$O(\varepsilon^0):\ f_i^{k,(0)} = f_i^{k,eq}, \tag{24}$$

$$O(\varepsilon^1):\ \left(\partial_{t_1} + \mathbf{e}_i\cdot\nabla_1\right) f_i^{k,(0)} = -\frac{1}{\tau\delta_t} f_i^{k,(1)} + (1-\frac{1}{2\tau})G_{1i}^k, \tag{25}$$

$$O(\varepsilon^2):\ \partial_{t_2} f_i^{k,(0)} + \left(\partial_{t_1} + \mathbf{e}_i\cdot\nabla_1\right) f_i^{k,(1)} + \frac{\delta_t}{2}\left(\partial_{t_1} + \mathbf{e}_i\cdot\nabla_1\right)^2 f_i^{k,(0)} = -\frac{1}{\tau\delta_t} f_i^{k,(2)}. \tag{26}$$

With the help of Eq. (25), Eq. (26) can be rewritten as

$$O(\varepsilon^2):\ \partial_{t_2} f_i^{k,(0)} + \left(\partial_{t_1} + \mathbf{e}_i\cdot\nabla_1\right)\left(1-\frac{1}{2\tau}\right) f_i^{k,(1)} + \frac{\delta_t}{2}\left(\partial_{t_1} + \mathbf{e}_i\cdot\nabla_1\right)\left(1-\frac{1}{2\tau}\right)G_{1i}^k = -\frac{1}{\tau\delta_t} f_i^{k,(2)}. \tag{27}$$

Taking the summations of Eqs. (25) and (27) and using $\sum_i f_i^{k,(n)} = \sum_i \mathbf{e}_i f_i^{k,(n)} = 0$ $(n = 1,2,\cdots)$ as well as Eq. (21), we can obtain

$$\partial_{t_1}\rho_k + \nabla_1\cdot(\rho_k\mathbf{u}) = 0, \tag{28}$$

$$\partial_{t_2}\rho_k = 0. \tag{29}$$

The continuity equation can be obtained by combining Eq. (28) with Eq. (29). Similarly, the first-order moments of Eqs. (25) and (27) yield, respectively

$$\partial_{t_1}(\rho_k\mathbf{u}) + \nabla_1\cdot(\rho_k\mathbf{uu}) = -\nabla_1 p_k, \tag{30}$$

$$\partial_{t_2}(\rho_k\mathbf{u}) + \nabla_1\cdot\left[\left(1-\frac{1}{2\tau}\right)\sum_i \mathbf{e}_i\mathbf{e}_i f_i^{k,(1)}\right] + \frac{\delta_t}{2}\nabla_1\cdot\left[\left(1-\frac{1}{2\tau}\right)\sum_i \mathbf{e}_i\mathbf{e}_i G_{1i}^k\right] = 0, \tag{31}$$

where $p_k = \rho_k (c_s^k)^2 = 0.5\rho_k c^2(1-\alpha_k)$ with $\phi_i^k$ being given by Eq. (13). Meanwhile, Eq. (25) gives

$$\sum_i \mathbf{e}_i\mathbf{e}_i f_i^{k,(1)} = -\tau\delta_t\left[\partial_{t_1}\left(\sum_i \mathbf{e}_i\mathbf{e}_i f_i^{k,(0)}\right) + \nabla_1\cdot\left(\sum_i \mathbf{e}_i\mathbf{e}_i\mathbf{e}_i f_i^{k,(0)}\right) - (1-\frac{1}{2\tau})\sum_i \mathbf{e}_i\mathbf{e}_i G_{1i}^k\right]. \tag{32}$$

Substituting Eq. (32) into Eq. (31) leads to

$$\partial_{t_2}\left(\rho_k \mathbf{u}\right) - \delta_t \nabla_1 \cdot \left\{ \left(\tau - 0.5\right) \left[ \partial_{t_1}\left(\sum_i \mathbf{e}_i \mathbf{e}_i f_i^{k,(0)}\right) + \nabla_1 \cdot \left(\sum_i \mathbf{e}_i \mathbf{e}_i \mathbf{e}_i f_i^{k,(0)}\right) - \sum_i \mathbf{e}_i \mathbf{e}_i G_{1i}^k \right] \right\}. \tag{33}$$

Using the new equilibrium distribution function given by Eq. (18), the off-diagonal elements of the third-order moment $\sum_i \mathbf{e}_i \mathbf{e}_i \mathbf{e}_i f_i^{k,(0)}$ can satisfy the required relationship, as shown in Eq. (19) (note that $f_i^{k,(0)} = f_i^{k,eq}$). Hence we can obtain the following constraints on the correction term $G_i^k$:

$$\sum_i e_{ix} e_{iy} G_i^k = \sum_i e_{ix} e_{iz} G_i^k = \sum_i e_{iy} e_{iz} G_i^k = 0\,. \tag{34}$$

However, the diagonal elements of the third-order moment $\sum_i \mathbf{e}_i \mathbf{e}_i \mathbf{e}_i f_i^{k,(0)}$ deviate from the required relationship. To remove the related error terms, the correction term $G_i^k$ should satisfy

$$\sum_i e_{ix}^2 G_i^k = \partial_x \left[\rho_k u_x \left(c^2 - 3(c_s^k)^2\right)\right], \tag{35}$$

$$\sum_i e_{iy}^2 G_i^k = \partial_y \left[\rho_k u_y \left(c^2 - 3(c_s^k)^2\right)\right], \tag{36}$$

$$\sum_i e_{iz}^2 G_i^k = \partial_z \left[\rho_k u_z \left(c^2 - 3(c_s^k)^2\right)\right]. \tag{37}$$

With these constraints, the error terms caused by the diagonal elements of the third-order moment of the equilibrium distribution function can be removed, and then the following equation can be derived from Eq. (33) by substituting Eqs. (16) and (19) as well as the above constraints into Eq. (33):

$$\partial_{t_2}\left(\rho_k \mathbf{u}\right) = \delta_t \nabla_1 \cdot \left[\left(\tau - 0.5\right) p_k \left(\nabla_1 \mathbf{u} + (\nabla_1 \mathbf{u})^{\mathrm{T}}\right)\right]. \tag{38}$$

Combining Eq. (38) with Eq. (30) through Eq. (23), the following macroscopic momentum equation can be obtained:

$$\partial_t\left(\rho_k \mathbf{u}\right) + \nabla \cdot \left(\rho_k \mathbf{u}\mathbf{u}\right) = -\nabla p_k + \nabla \cdot \left[\rho_k \nu_k \left(\nabla \mathbf{u} + (\nabla \mathbf{u})^{\mathrm{T}}\right)\right], \tag{39}$$

where the kinematic viscosity $\nu_k$ is given by $\nu_k = \left(\tau - 0.5\right)\delta_t (c_s^k)^2$.

To sum up, the new equilibrium distribution function given by Eq. (18) and the correction term in Eq. (20) constitute the improvements for removing the error terms in Eq. (14). The form of the correction term can be determined by the aforementioned constraints. Particularly, since the constraints are given in the form of the moments of $G_i^k$, the correction term can be readily obtained in the moment space, namely $\mathbf{C}^k = \mathbf{M}\mathbf{G}^k$, where $\mathbf{C}^k$ is the correction term in the moment space and $\mathbf{M}$ the transformation matrix of an MRT collision operator. Considering such a feature of the MRT collision operator and its better numerical stability over the BGK collision operator, in what follows we shall construct the improved three-dimensional color-gradient LB model based on the MRT collision operator.

### B. Improved model based on the MRT collision operator

Using the MRT collision operator [35-38], the single-phase collision operator with the correction term can be written as follows:

$$(\Omega_i^k)^{(1)} = -\Lambda_{ij}\left[f_j^k(\mathbf{x},t) - f_j^{k,eq}(\mathbf{x},t)\right] + \delta_t\left(\delta_{ij} - \frac{1}{2}\Lambda_{ij}\right)G_j(\mathbf{x},t), \tag{40}$$

where $\delta_{ij}$ is the Kronecker delta and $\Lambda_{ij} = \left(\mathbf{M}^{-1}\mathbf{S}\mathbf{M}\right)_{ij}$, in which $\mathbf{M}$ is the transformation matrix and $\mathbf{S}$ is a diagonal matrix for the relaxation times. Obviously, when $\Lambda_{ij} = \delta_{ij}/\tau$ (i.e., the BGK collision operator), Eq. (40) reduces to Eq. (20). Through the transformation matrix, the right-hand side of Eq. (40) can be implemented in the moment space:

$$\bar{\mathbf{m}}^k = -\mathbf{S}\left(\mathbf{m}^k - \mathbf{m}^{k,eq}\right) + \delta_t\left(\mathbf{I} - \frac{\mathbf{S}}{2}\right)\mathbf{C}^k, \tag{41}$$

where $\mathbf{I}$ is the unit matrix, $\mathbf{m}^k = \mathbf{M}\mathbf{f}^k$ is the related moments, $\mathbf{m}^{k,eq}$ is the equilibria in the moment space, and $\mathbf{C}^k = \mathbf{M}\mathbf{G}^k$ is the correction term in the moment space. The equilibria can be obtained through $\mathbf{m}^{k,eq} = \mathbf{M}\mathbf{f}^{k,eq}$ with $\mathbf{f}^{k,eq} = (f_0^{k,eq}, f_1^{k,eq}, \ldots, f_{18}^{k,eq})^{\mathrm{T}}$ being defined by Eq. (18), i.e., the new equilibrium distribution function. Correspondingly, Eq. (40) can be rewritten as

$$(\Omega_i^k)^{(1)} = \left(\mathbf{M}^{-1}\right)_{ij}\left(\bar{\mathbf{m}}^k\right)_j. \tag{42}$$

The recoloring steps for the red and blue fluids are still given by Eqs. (10) and (11), respectively.

In the present work, the improved color-gradient MRT-LB model is constructed using the D3Q19 lattice and the following transformation matrix $\mathbf{M}$ is employed [39,40]:

$$\mathbf{M}=\begin{bmatrix} \langle 1| \\ \langle e_{ix}| \\ \langle e_{iy}| \\ \langle e_{iz}| \\ \langle |\mathbf{e}_i|^2| \\ \langle (2e_{ix}^2-e_{iy}^2-e_{iz}^2)| \\ \langle (e_{iy}^2-e_{iz}^2)| \\ \langle e_{ix}e_{iy}| \\ \langle e_{ix}e_{iz}| \\ \langle e_{iy}e_{iz}| \\ \langle e_{ix}^2e_{iy}| \\ \langle e_{ix}e_{iy}^2| \\ \langle e_{ix}^2e_{iz}| \\ \langle e_{ix}e_{iz}^2| \\ \langle e_{iy}^2e_{iz}| \\ \langle e_{iy}e_{iz}^2| \\ \langle e_{ix}^2e_{iy}^2| \\ \langle e_{ix}^2e_{iz}^2| \\ \langle e_{iy}^2e_{iz}^2| \end{bmatrix}, \tag{43}$$

where $|\mathbf{e}_i|^2 = e_{ix}^2 + e_{iy}^2 + e_{iz}^2$. The first ten vectors are related to the macroscopic density, momentum, and the viscous stress tensor, whereas the other vectors are related to high-order moments that do not affect the Navier-Stokes level hydrodynamics. The detailed form of the transformation matrix $\mathbf{M}$ is given in the Appendix. The relaxation matrix $\mathbf{S}$ in Eq. (41) is defined as [39,40]

$$\mathbf{S} = \mathrm{diag}\left(1, 1, 1, 1, \tau_e^{-1}, \tau_v^{-1}, \tau_v^{-1}, \tau_v^{-1}, \tau_v^{-1}, \tau_v^{-1}, \tau_q^{-1}, \tau_q^{-1}, \tau_q^{-1}, \tau_q^{-1}, \tau_q^{-1}, \tau_q^{-1}, \tau_\pi^{-1}, \tau_\pi^{-1}, \tau_\pi^{-1}\right), \tag{44}$$

where the relaxation times $\tau_v$ and $\tau_e$ determine the shear and bulk viscosities, respectively, while $\tau_q$ and $\tau_\pi$ are related to non-hydrodynamic moments. The equilibria in the moment space can be obtained by $\mathbf{m}^{k,eq} = \mathbf{M}\mathbf{f}^{k,eq}$ (see the Appendix for details) and the correction term in the moment space can be derived from $\mathbf{C}^k = \mathbf{M}\mathbf{G}^k$. According to the moment set in Eq. (43) and the constraints on $G_i^k$ (see Eq. (21) and Eqs. (34)-(37)), the following correction term can be obtained in the moment space:

$$
\mathbf{C}^k = \mathbf{M}\mathbf{G}^k = \begin{bmatrix} 0 \\ 0 \\ 0 \\ 0 \\ Q_x + Q_y + Q_z \\ 2Q_x - Q_y - Q_z \\ Q_y - Q_z \\ 0 \\ 0 \\ 0 \\ 0 \\ 0 \\ 0 \\ 0 \\ 0 \\ 0 \\ 0 \\ 0 \\ 0 \end{bmatrix}, \tag{45}
$$

where $Q_x$, $Q_y$, and $Q_z$ are given by (see also Eqs. (35)-(37)), respectively

$$
Q_x = \partial_x \left[ \rho_k u_x \left( c^2 - 3(c_s^k)^2 \right) \right], \tag{46}
$$

$$
Q_y = \partial_y \left[ \rho_k u_y \left( c^2 - 3(c_s^k)^2 \right) \right], \tag{47}
$$

$$
Q_z = \partial_z \left[ \rho_k u_z \left( c^2 - 3(c_s^k)^2 \right) \right], \tag{48}
$$

where $(c_s^k)^2 = 0.5c^2 \left(1 - \alpha_k\right)$. The high-order moments of $G_i^k$ have been set to zero in deriving Eq. (45). In numerical implementation, the second-order isotropic difference scheme is applied to the spatial gradients in Eqs. (46)-(48), i.e.,

$$
Q_\alpha \approx \frac{3}{c^2 \delta_t} \sum_i \omega_i \bar{Q}_\alpha \left( \mathbf{x} + \mathbf{e}_i \delta_t \right) e_{i\alpha}, \tag{49}
$$

where $\alpha$ denotes the *x*, *y*, or *z* coordinate and $\bar{Q}_\alpha = \rho_k u_\alpha \left( c^2 - 3(c_s^k)^2 \right)$. It can be found that such a calculation is the same as the calculation of the color gradient $\nabla \rho^N$ in Eq. (8).

The Chapman-Enskog analysis can also be applied to the MRT collision operator, which is similar to that of the BGK collision operator. Readers are referred to Refs. [40-42] about the Chapman-Enskog analysis of the three-dimensional MRT-LB method. It can be found that, using the equilibria $\mathbf{m}^{k,eq}$ in the Appendix and the correction term $\mathbf{C}^k$ given by Eq. (45), the following macroscopic momentum equation can be derived in the low Mach number limit:

$$\partial_t(\rho_k \mathbf{u}) + \nabla\cdot(\rho_k \mathbf{uu}) = -\nabla p_k + \nabla\cdot\left[\mu_k\left(\nabla\mathbf{u} + (\nabla\mathbf{u})^{\mathrm{T}}\right) - \frac{2}{3}\mu_k(\nabla\cdot\mathbf{u})\mathbf{I} + \mu_k^b(\nabla\cdot\mathbf{u})\mathbf{I}\right], \tag{50}$$

where the dynamic shear viscosity $\mu_k$ and the bulk viscosity $\mu_k^b$ are given by, respectively

$$\mu_k = \rho_k (c_s^k)^2\left(\tau_v - \frac{1}{2}\right)\delta_t, \quad \mu_k^b = \frac{2}{3}\rho_k (c_s^k)^2\left(\tau_e - \frac{1}{2}\right)\delta_t. \tag{51}$$

The kinematic viscosity $\nu_k$ is given by $\nu_k = \mu_k/\rho_k$. When $\tau_e = \tau_v$, Eq. (50) reduces to Eq. (39).

To ensure the smoothness of the relaxation time $\tau_v$ (corresponding to $\tau$ in the BGK collision model) across the interface, $\tau_v$ is calculated as follows [21,23]:

$$\tau_v = \begin{cases} \tau_v^R, & \rho^N > \delta, \\ g^R(\rho^N), & \delta \ge \rho^N > 0, \\ g^B(\rho^N), & 0 \ge \rho^N \ge -\delta, \\ \tau_v^B, & \rho^N < -\delta, \end{cases} \tag{52}$$

where $\delta$ is a free parameter related to the interface thickness and is usually set as $\delta = 0.98$ [12], and $g^R$ and $g^B$ are parabolic functions of $\rho^N$ (its definition is given below Eq. (8)), as shown in Refs. [21,23]. The relaxation times $\tau_v^R$ and $\tau_v^B$ in Eq. (52) are determined by the kinematic viscosities of the red and blue fluids, i.e., $\nu_R = (c_s^R)^2\left(\tau_v^R - 0.5\right)\delta_t$ and $\nu_B = (c_s^B)^2\left(\tau_v^B - 0.5\right)\delta_t$, respectively.

The surface tension in Eq. (9) depends on the relaxation time. A simple treatment to make the surface tension independent of the relaxation time is to change the perturbation operator from Eq. (8) to $(\Omega_i^k)^{(2),\mathrm{new}} = \frac{1}{\tau}(\Omega_i^k)^{(2)}$, and then the surface tension is given by $\sigma = 2(A_R + A_B)c^4\delta_t/9$. Correspondingly, the perturbation operator within the framework the MRT-LB method can be redefined as

$$(\Omega_i^k)^{(2),\mathrm{new}} = \Lambda_{ij}(\Omega_j^k)^{(2)}. \tag{53}$$

Similar to the single-phase collision operator in Eq. (40), the perturbation operator given by Eq. (53) can also be executed in the moment space.

## IV. Numerical results and discussion

In this section, numerical simulations are carried out to validate the improved three-dimensional color-gradient LB model. First, the test of a moving droplet in a uniform flow field is employed to verify the Galilean invariance of the improved model. Subsequently, the numerical accuracy of the improved model is demonstrated through simulating the layered two-phase flow and three-dimensional Rayleigh-Taylor instability. Finally, the capability of the improved model for simulating dynamic multiphase flows at a large density ratio is validated by the simulation of droplet impact on a solid wall.

### A. Moving droplet in a uniform flow field

In LB community [43,44], it has been reported that a circular droplet in a uniform flow field will become an elliptic one when employing a multiphase LB model with broken Galilean invariance. To verify the Galilean invariance of the proposed improved color-gradient LB model, the test of a moving circular droplet in a uniform flow field is considered. Our simulations are carried out in a domain divided into $N_x \times N_y \times N_z = 140 \times 140 \times 4$ lattices. A circular droplet of radius $r_0 = 30$ (lattice unit) is placed at the center of the computational domain and brought to the equilibrium state at rest. Then the two parallel plates in the $y$ direction begin to move with a constant velocity $U = 0.02$ at $t = 0$. The Zou-He boundary scheme [45] is applied in the $y$ direction and the periodic boundary condition is employed in the $x$ and $z$ directions.

The initial densities of the red and blue fluids are taken as $\rho_R^{\text{in}} = 3$ and $\rho_B^{\text{in}} = 1$, respectively, with $\alpha_R$ and $\alpha_B$ being set to 0.9 and 0.7, respectively, which satisfy $\rho_R^{\text{in}} / \rho_B^{\text{in}} = (1-\alpha_B)/(1-\alpha_R)$ [26]. The parameters $A_R$ and $A_B$ for the surface tension are $A_R = A_B = 0.01$ and the parameter $\beta$ in Eqs. (10) and (11) is chosen as $\beta = 0.5$. The relaxation time $\tau_v$ is determined by Eq. (52) and the other relaxation times are set to 1.0. The dynamics viscosities are chosen as $\mu_R = \mu_B = 0.075$. Figure 1 shows the simulated snapshots of a moving circular droplet. For comparison, the numerical results of the color-gradient MRT-LB model without the improvements are also presented, which is the MRT version of the three-dimensional color-gradient LB model of Liu *et al.* [24] and is hereinafter referred to as the original model. When the original model is employed, the shape of the droplet becomes elliptic, as shown in Fig. 1(a), which means that the lack of Galilean invariance leads to deformation of the droplet. On the contrary, from Fig. 1(b) we can see that the improved color-gradient model allows the droplet to retain its circular shape, demonstrating that the Galilean invariance is restored in the improved model.

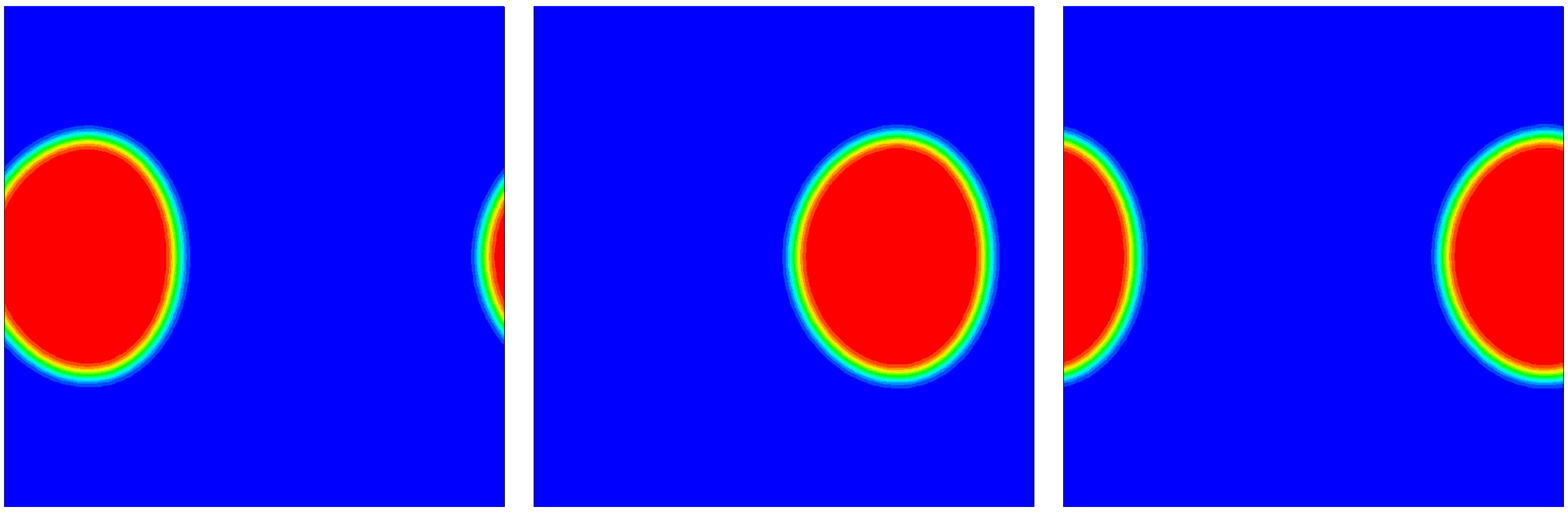

(a) original model

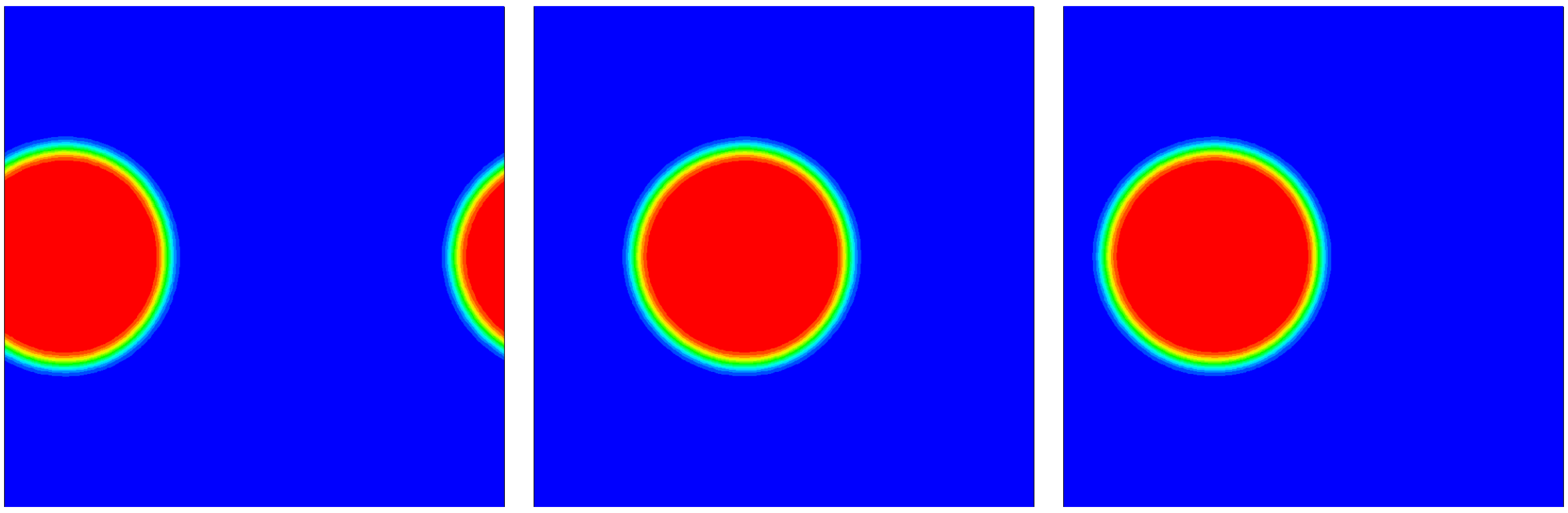

(b) improved model

**FIG. 1**. Density contours of a moving droplet simulated by (a) the original model and (b) the improved model. From left to right: $t = 60000\delta_t$, $80000\delta_t$, and $95000\delta_t$.

### B. Layered two-phase flow in a channel

In this subsection, the layered two-phase flow between two parallel plates is simulated to validate the numerical accuracy of the improved color-gradient LB model. As shown in Fig. 2, the channel height is $h = 2b$ in the $y$ direction with $y = 0$ at the center of the channel. The red fluid is initially located in the central region $-a \le y \le a$, whereas the blue fluid is located in the regions $a < |y| \le b$. The layered two-phase flow is driven by a constant body force $(G,\ 0,\ 0)$. By assuming a Poiseuille-type flow in the channel, we can obtain the following the analytical solution for the velocity profile [12]:

$$u_x(y)=\begin{cases} A_1 y^2 + C_1 & (0 \le |y| \le a), \\ A_2 y^2 + B_2 y + C_2 & (a \le |y| \le b), \end{cases} \tag{54}$$

where the coefficients are defined as

$$A_1 = -\frac{G}{2\rho_R \nu_R},\quad A_2 = -\frac{G}{2\rho_B \nu_B},\quad B_2 = 2(A_1 M - A_2)a,$$

$$C_1 = (A_2 - A_1)a^2 - B_2(b-a) - A_2 b^2,\quad C_2 = -A_2 b^2 - B_2 b, \tag{55}$$

in which $M = \mu_R / \mu_B$ is the dynamic viscosity ratio [12].

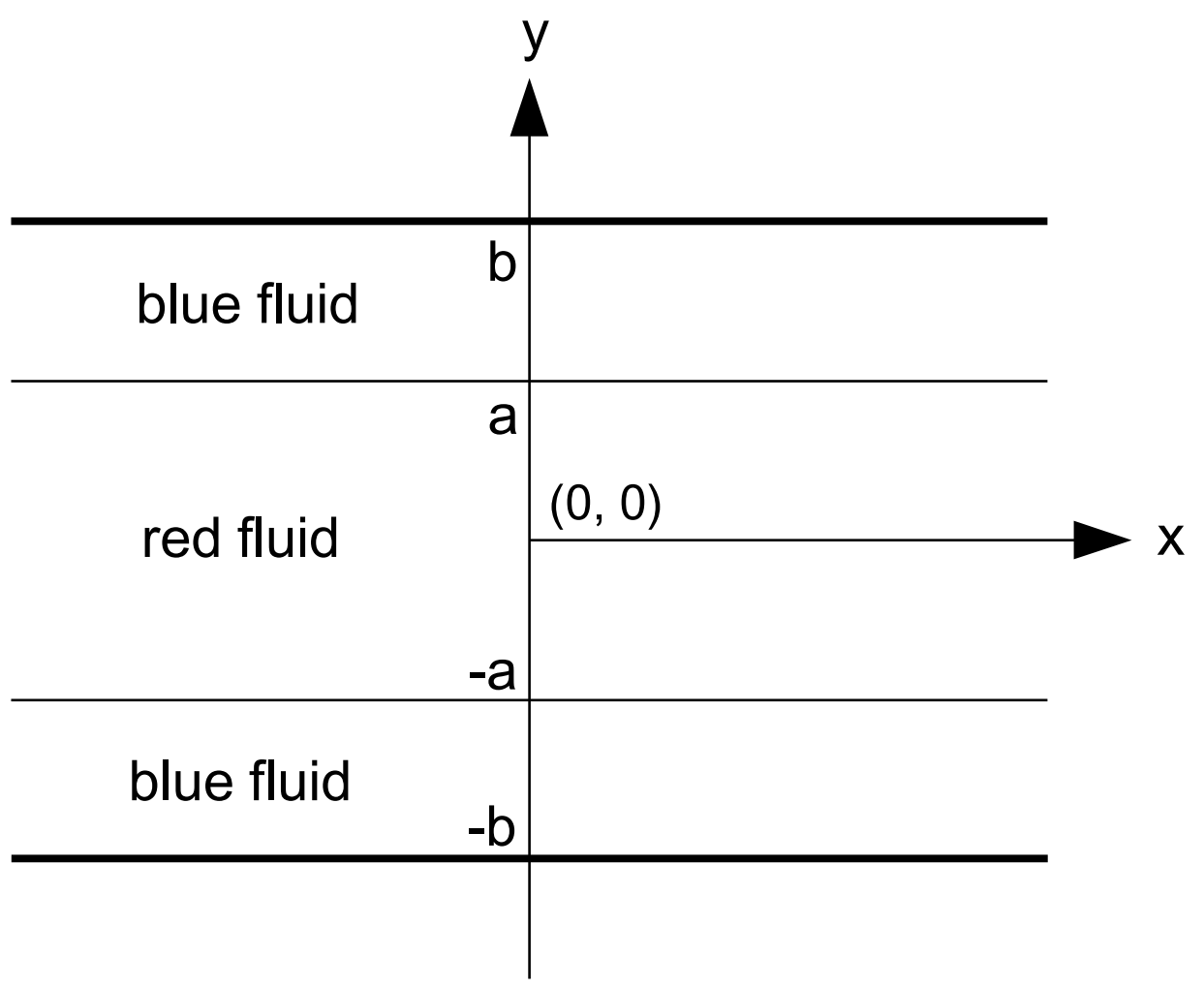

**FIG. 2**. Schematic of the layered two-phase flow between two parallel plates.

In our simulations, the computational domain is divided into $N_x \times N_y \times N_z = 10 \times 100 \times 4$ lattices with $a = 25$ and $b = 50$. The non-slip boundary condition [45] is applied to the two parallel plates, while the periodic boundary condition is employed in the *x* and *z* directions. Three cases are investigated:

Case A: $\rho_R^{\text{in}} = 0.1$, $\rho_B^{\text{in}} = 0.8$, $\alpha_B = 0.9$, $M = 1/8$;

Case B: $\rho_R^{\text{in}} = 0.8$, $\rho_B^{\text{in}} = 0.1$, $\alpha_B = 0.2$, $M = 8$;

Case C: $\rho_R^{\text{in}} = 0.008$, $\rho_B^{\text{in}} = 8$, $\alpha_B = 0.9992$, $M = 1/40$. The parameter $\alpha_R$ is determined via $\rho_R^{\text{in}} / \rho_B^{\text{in}} = (1-\alpha_B)/(1-\alpha_R)$. The parameters $A_R$ and $A_B$ for the surface tension are chosen as $A_R = A_B = 0.0001$. The constant body force in the *x* direction is taken as $G = 1.5 \times 10^{-8}$.

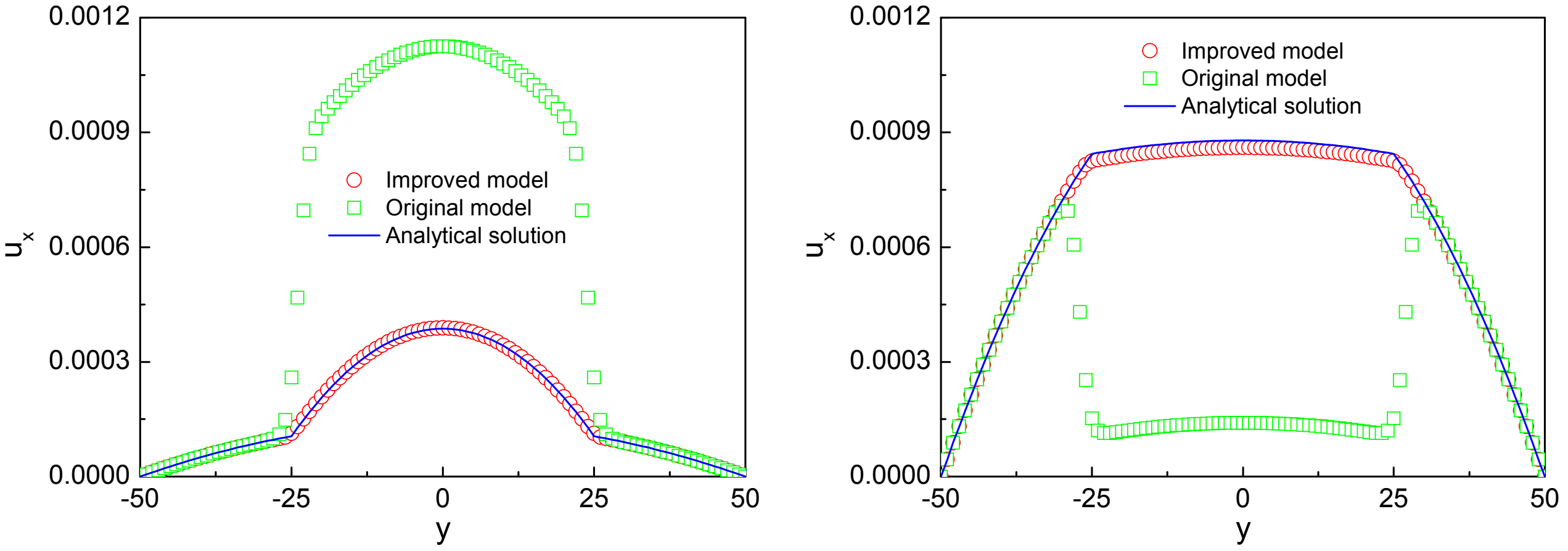

(a) Case A (b) Case B

**FIG. 3**. Simulation of layered two-phase flow in a channel. Comparison of the velocity profiles obtained by the original and improved color-gradient LB models for cases A and B.

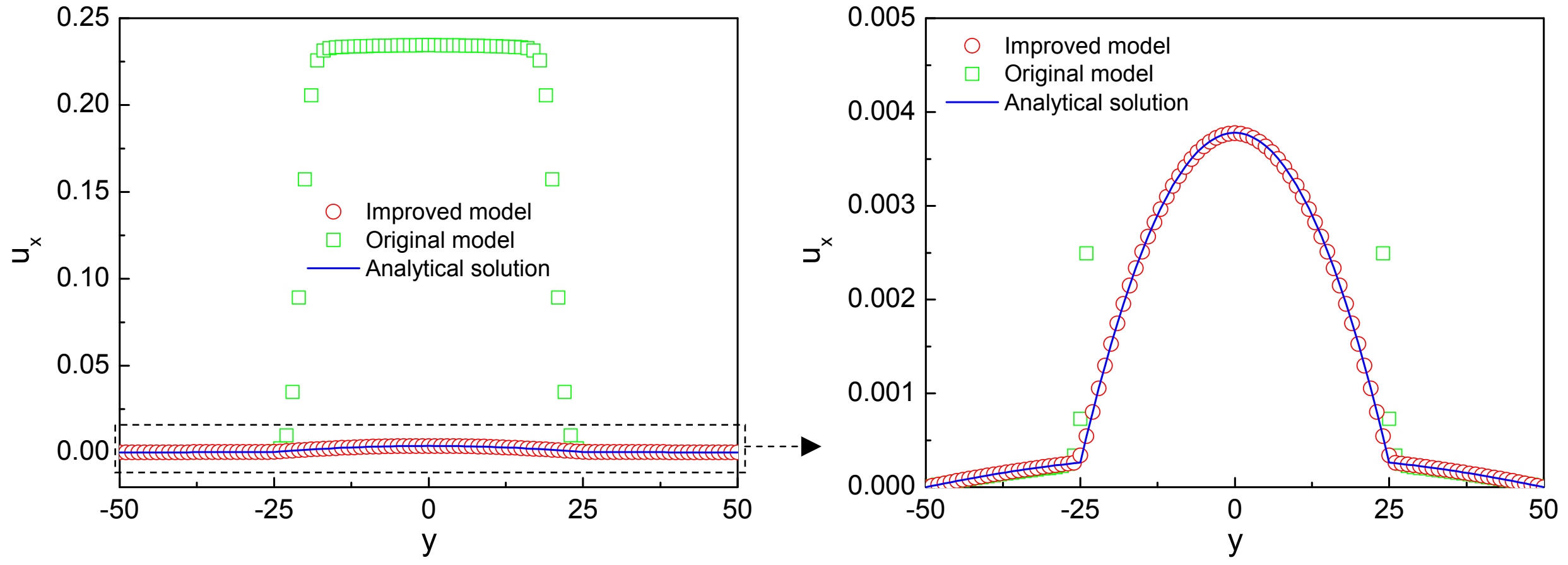

**FIG. 4**. Simulation of layered two-phase flow in a channel. Comparison of the velocity profiles obtained by the original and improved color-gradient LB models for case C.

Figures 3 and 4 display the numerical results obtained by the original and improved color-gradient LB models for cases A, B, and C. For comparison, the analytical solutions are also presented. From the figures we can see that the velocity profiles predicted by the improved model are in good agreement with the analytical solutions, whereas the numerical results obtained by the original model significantly deviate from the analytical ones. Particularly, it can be seen that the main numerical errors appear within the interval $y \in [-25, 25]$. This is because the error terms in Eq. (14) cannot be neglected due to the abrupt change of the momentum across the interfaces around $y = \pm 25$. To quantify the numerical simulations, the relative error between the numerical results and the analytical solutions is evaluated, which is defined as $E_u = \sum_y \left| u_x(y) - u_x^a(y) \right| / \sum_y \left| u_x^a(y) \right|$, where $u_x^a(y)$ denotes the analytical solution. For the improved model, the relative errors of cases A, B, and C are $E_u = 0.66\%$, $1.42\%$, and $0.36\%$, respectively, while the relative errors yielded by the original model for the three cases are $E_u = 44.6\%$, 59.7%, and 6904%, respectively. Here it can be seen that the error of case C caused by the original model is much larger than those of cases A and B, which is attributed to the fact that the ratio $\rho_B^{\text{in}} / \rho_R^{\text{in}}$ is very large in case C, i.e., $\rho_B^{\text{in}} / \rho_R^{\text{in}} = 8/0.008 = 1000$.

### C. Three-dimensional Rayleigh-Taylor instability

The phenomenon of Rayleigh-Taylor instability is associated with the penetration of a heavy fluid into a light fluid and can be found in a wide range of scientific and environmental fields. This problem involves complex interfacial interactions and has been intensively studied because of its practical and scientific importance [46-48]. It consists of two layers of fluid at rest: a heavy fluid is on top of a light fluid. The heavy fluid accelerates into the light fluid under the action of the gravity. In the present study, the test of three-dimensional Rayleigh-Taylor instability is employed to investigate the capability of the improved model for modeling multiphase flows with complex interfacial interactions. The computational domain is a rectangular box of $\Omega = [0, L] \times [0, L] \times [0, 4L]$. The non-slip boundary condition is applied to the upper and lower solid walls, while the periodic boundary condition is employed at the four vertical boundaries.

In our simulations, the red (heavy) fluid is placed above the blue (light) fluid and the Atwood number $\text{At} = \left( \rho_R^{\text{in}} - \rho_B^{\text{in}} \right) / \left( \rho_R^{\text{in}} + \rho_B^{\text{in}} \right)$ is set to 0.5 for the sake of comparing our numerical results with those reported in the literature [48]. The Reynolds number is defined as $\text{Re} = L\sqrt{Lg}/\nu$, where $g$ is the gravitational acceleration and $\nu$ is the kinematic viscosity. In this problem, the kinematic viscosities of the two fluids are identical. The characteristic velocity of the system is taken as $U = \sqrt{Lg} = 0.04$, $L$ is chosen as $L = 125$ (lattice unit), and the Reynolds number is set to $\mathrm{Re} = 1024$. The investigated Rayleigh-Taylor instability develops from the following single mode initial perturbation:

$$\frac{h(x, y)}{L} = 0.05\left[\cos\left(\frac{2\pi x}{L}\right) + \cos\left(\frac{2\pi y}{L}\right)\right], \tag{56}$$

where $h$ is the height of the fluid interface.

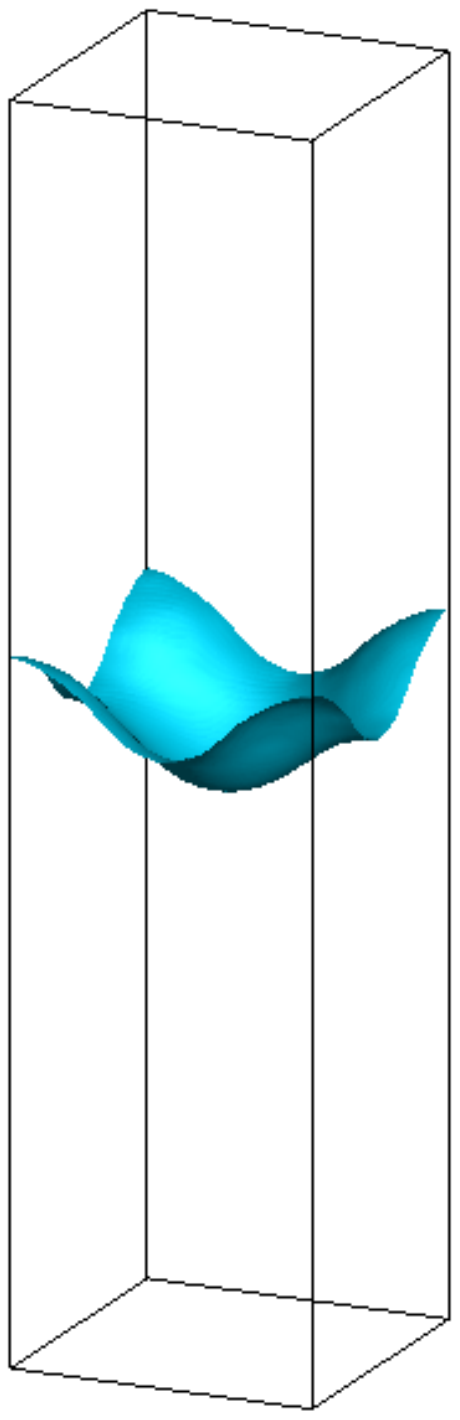

(a) original model

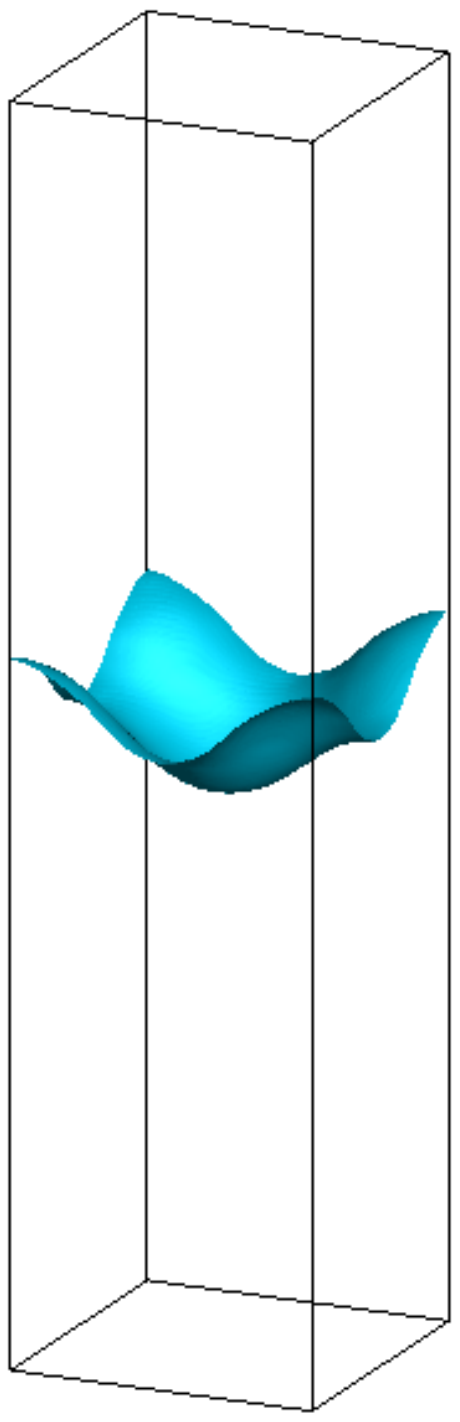

(b) improved model

**FIG. 5**. Simulation of three-dimensional Rayleigh-Taylor instability. Snapshots of the fluid interface obtained by (a) the original model and (b) the improved model at $t^* = 1$, 2, and 3 (from left to right).

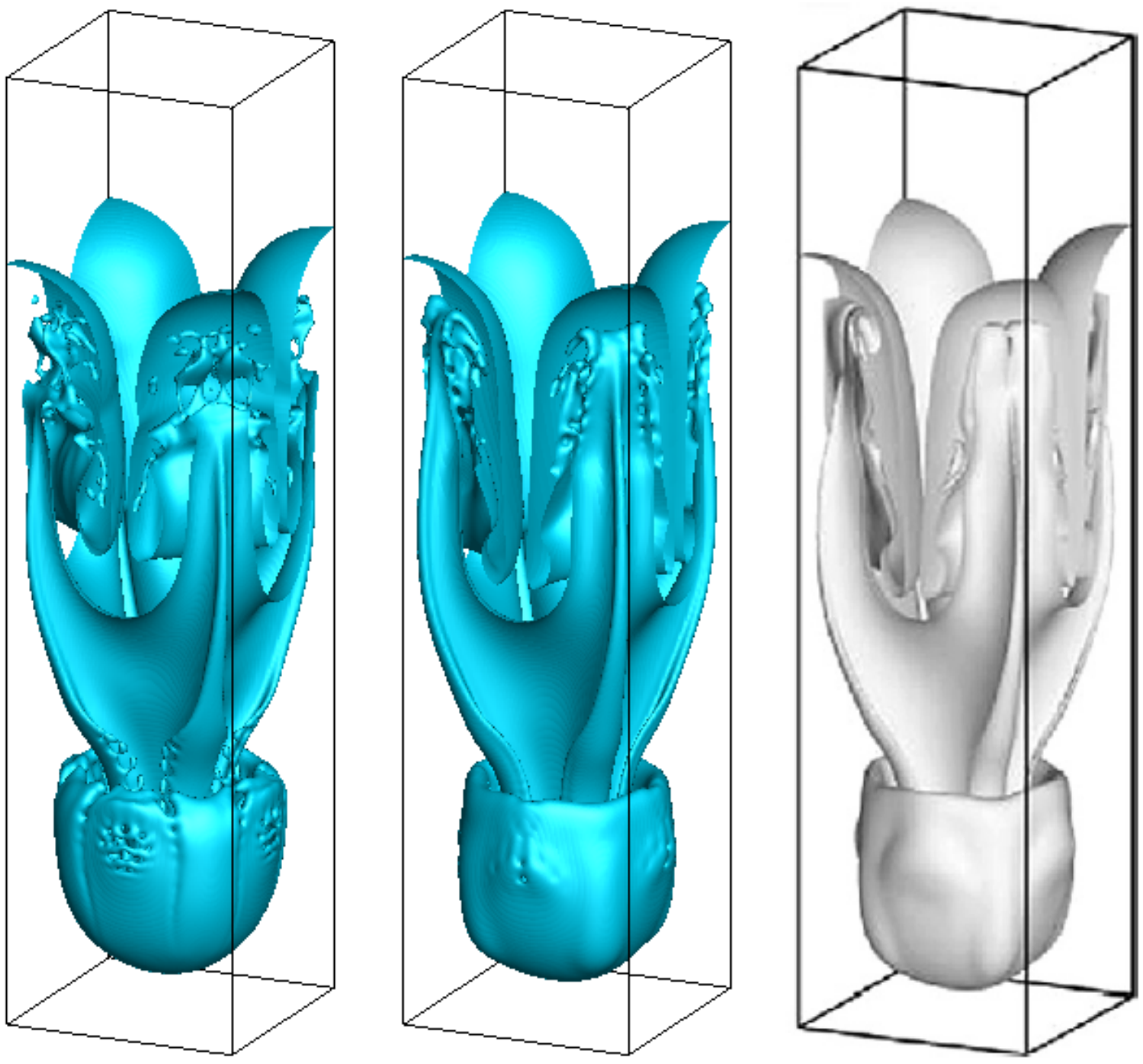

**FIG. 6**. Comparison of the fluid interface at $t^* = 4$ obtained by (a) the original color-gradient model, (b) the improved color-gradient model, and (c) a multiphase flux solver in Ref. [48].

Figure 5 shows the evolution of the fluid interface simulated by the original and improved color-gradient models at $t^* = 1$, 2, and 3, where the time $t^*$ is non-dimensional and is normalized by the reference time $t_{\mathrm{ref}} = \sqrt{L/g}$. From the numerical results of both models we can see that the heavy and light fluids penetrate into each other as time increases. Specifically, at $t^* = 1$ it can be seen that a spike is formed in the middle due to the downward movement of the heavy fluid and bubbles are formed on the sides because of the rising of light fluid. At the early stage, the numerical results of the two models show the same interface shapes. Subsequently, the first roll-up of the heavy fluid appears in the neighborhood of the saddle points, as can be seen at $t^* = 2$, and we can find that the shapes of the bubbles become a little different for the original and improved models. Later, at $t^* = 3$ the second roll-up takes place at the edge of the spike. At this time, the shapes of the spike obtained by the original and improved models are quite different, which is attributed to the fact that the downward velocity of the spike gradually increases and the error terms in Eq. (14) accordingly becomes non-negligible.

As time goes by, significant differences can be observed between the results of the improved model and those of the original model, as shown in Fig. 6, which shows a comparison of the fluid interface at $t^* = 4$ obtained by the original model, the improved model, and a multiphase flux solver in Ref. [48]. At this stage, two extra layers of the heavy fluid are folded upward as a result of the stretch of the two roll-ups: one forms a skirt around the spike and the other forms a girdle inside the bubble. Particularly, it can be seen that the shapes of the spike and girdle predicted by the original model obviously deviate from those obtained by a multiphase flux solver [48], while the shapes simulated by the improved model are in good agreement with those reported in Ref. [48]. Figure 7 depicts the evolution of the interface positions of the bubble front, the spike tip, and the saddle point. As shown in the figure, the interface positions predicted by our improved model agree well with the results of Wang *et al.* [48].

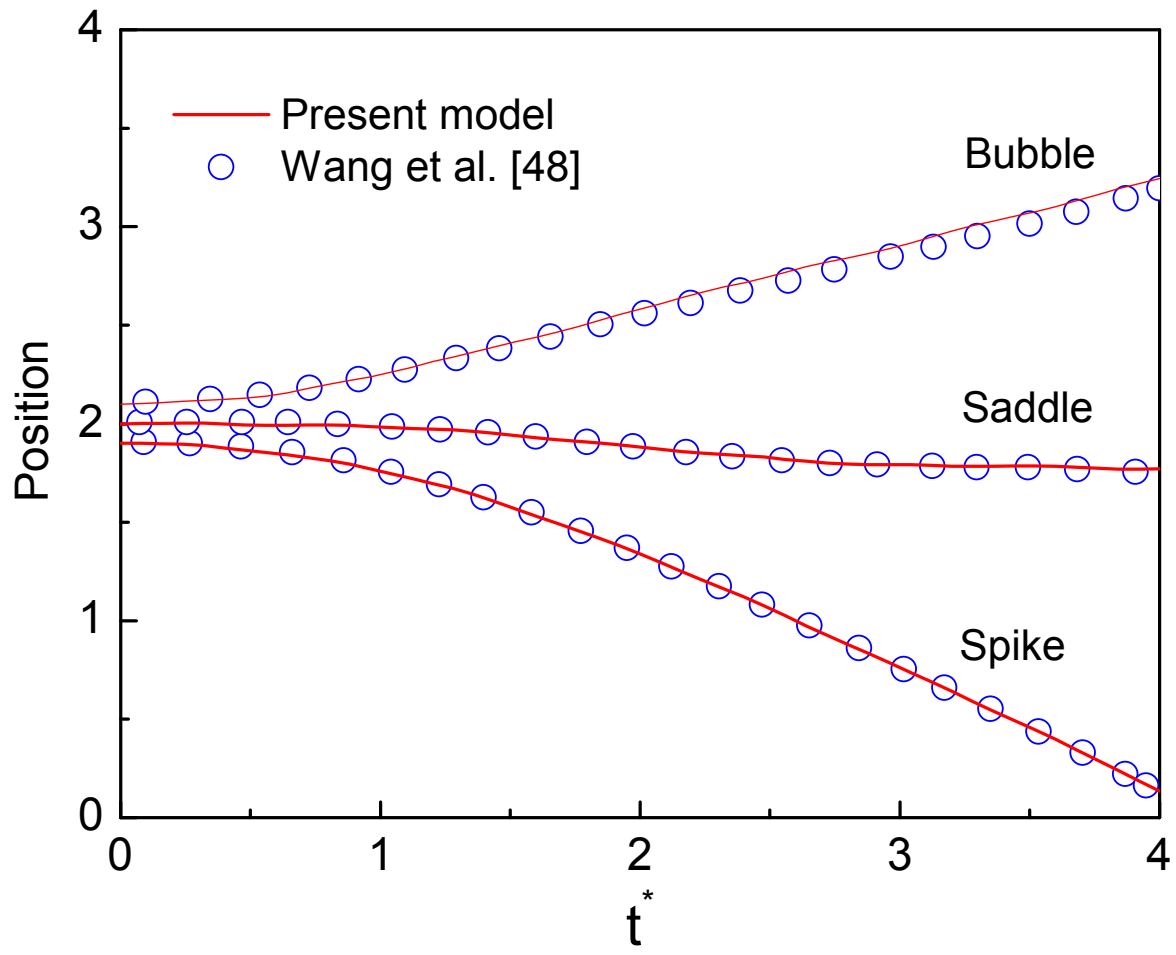

**FIG. 7**. The positions of the bubble front, the spike tip, and the saddle point versus time. The Atwood number is 0.5 and the Reynolds number is 1024.

### D. Droplet impact on a solid surface

Finally, the impingement of a droplet on a flat surface is simulated to validate the capability of the proposed color-gradient MRT-LB model for simulating dynamic multiphase flows at a relatively large density ratio. Impingement of droplets on a solid surface is a very important phenomenon in many engineering applications, such as ink-jet printing and spray cooling. The dynamics of droplet impact on solid surfaces is usually governed by the following two non-dimensional parameters:

$$\mathrm{Re} = \frac{\rho_R U_0 D_0}{\mu_R}, \quad \mathrm{We} = \frac{\rho_R U_0^2 D_0}{\sigma}, \tag{57}$$

where $\mathrm{Re}$ and $\mathrm{We}$ are the Reynolds number and the Weber number, respectively. In Eq. (57), $U_0$ is the impact speed of the droplet and $D_0$ is the initial diameter of the droplet.

The computational domain is divided into $N_x \times N_y \times N_z = 300 \times 300 \times 150$ lattices. The non-slip boundary condition is employed at the solid surface and the periodic condition is applied in the *x* and *y* directions. Initially, a spherical droplet of diameter $D_0 = 100$ (lattice unit) is placed on the center of the bottom flat surface. The initial velocity of the droplet is given by $\mathbf{u}_0 = (u_x, u_y, u_z) = (0, 0, -U_0)$, in which $U_0 = 0.006$. The initial densities of the red and blue fluids are given by $\rho_R^{\mathrm{in}} = 8$ and $\rho_B^{\mathrm{in}} = 0.08$, respectively, with $\alpha_B = 0.2$. The equilibrium contact angle of the flat surface is taken as $\theta \approx 90^{\circ}$ and the parameters $A_R$ and $A_B$ for the surface tension are chosen as $A_R = A_B = 0.00225$, which leads to the surface tension $\sigma \approx 0.001$. Correspondingly, the Weber number is $\mathrm{We} \approx 28.8$.

In our simulations, the Reynolds number varies from $\mathrm{Re} = 75$ to $1000$. Figure 8 displays some snapshots of the droplet impingement process at $\mathrm{Re} = 1000$. As shown in the figure, immediately after the impingement, the shape of the droplet resembles a truncated sphere (Fig. 8a). Later, a lamella is formed as the liquid moves radially outwards (Figs. 8b and 8c). The lamella continues to grow radially (Fig. 8d) until the maximum spreading diameter is reached and the spreading process ends, during which the kinetic energy is transformed into the surface energy by increasing the area of the droplet [48]. After reaching the maximum spreading diameter, the lamella begins to retract because of the surface tension, as can be seen in Figs. 8(e) and 8(f). These observations agree well with those reported in the previous studies [49-52].

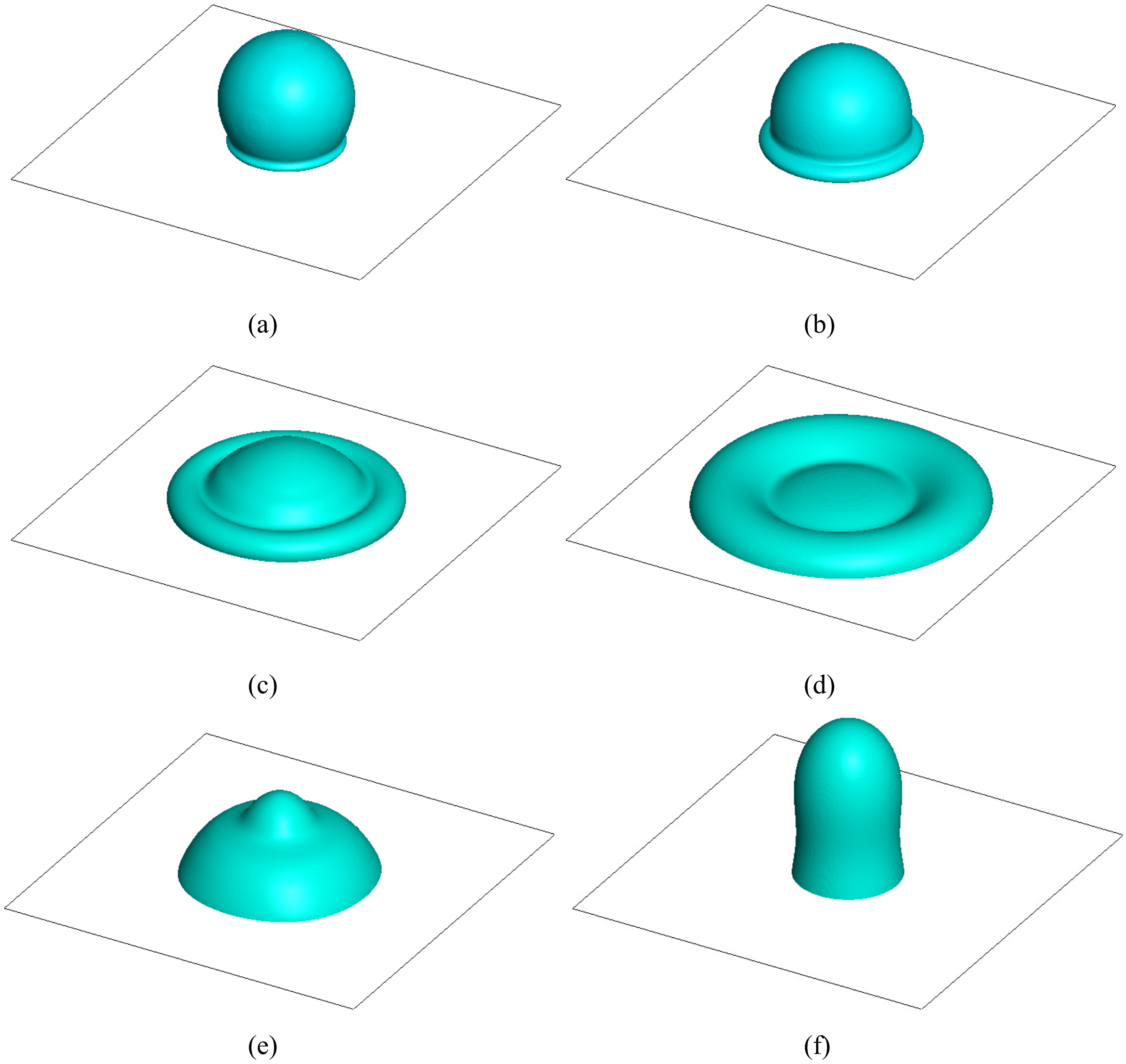

**FIG. 8**. Snapshots of droplet impact on a flat surface at $\mathrm{Re}=1000$ with $\rho_R/\rho_B=100$. (a) $t=2000\delta_t$, (b) $t=4000\delta_t$, (c) $t=10000\delta_t$, (d) $t=30000\delta_t$, (e) $t=60000\delta_t$, and (f) $t=90000\delta_t$.

To quantify the numerical results, the maximum spreading factor $D_{\max}/D_0$ obtained by the proposed color-gradient LB model is compared with the data reported in the literature. In Ref [50], Asai *et al.* established a correlation formula for the maximum spreading factor based on their experimental data: $D_{max}/D_0 = 1 + 0.48\mathrm{We}^{0.5}\exp\left(-1.48\mathrm{We}^{0.22}\,\mathrm{Re}^{-0.21}\right)$. In addition, Scheller and Bousfield [51] also proposed a correlation formula by plotting their experimental data versus $\mathrm{Oh\,Re}^2 = \sqrt{\mathrm{We}}\,\mathrm{Re}$, in which $\mathrm{Oh} = \sqrt{\mathrm{We}}/\mathrm{Re} = \mu_R/\sqrt{\rho_R\sigma D_0}$ is the Ohnesorge number. Figure 9 shows a comparison of the maximum spreading factor between the experimental correlation formula of Asai *et al.* [50], the experimental data of Scheller and Bousfield [51], and the numerical results predicted by the proposed color-gradient LB model. From the figure it can be seen that our numerical results are in good agreement with the experimental correlation/data reported in the previous studies, demonstrating that the improved color-gradient LB model is capable of simulating dynamic multiphase flows at a relatively large density ratio.

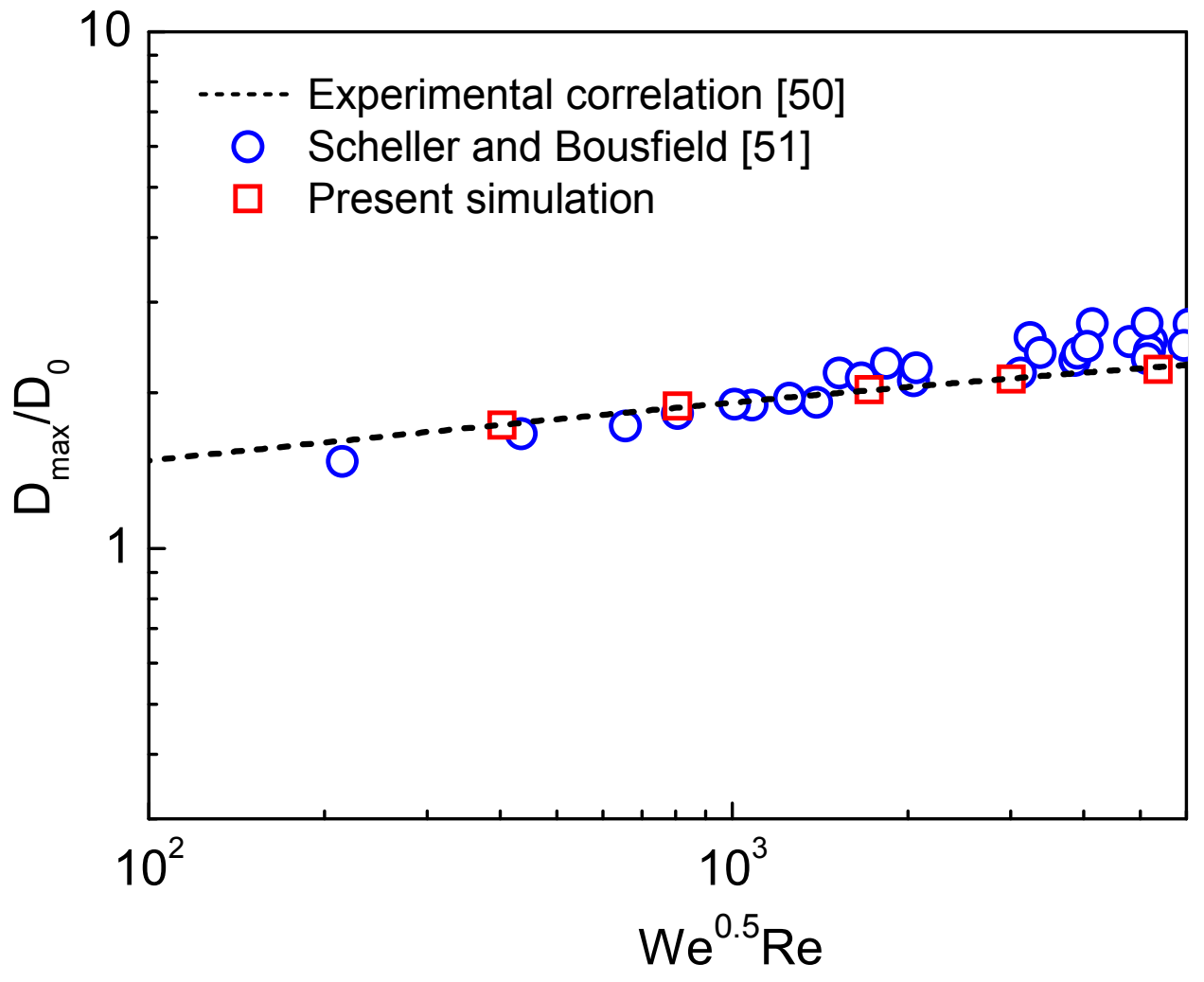

**FIG. 9**. Comparison of the maximum spreading factor between the present numerical results, the experimental correlation in Ref. [50], and the experimental data in Ref. [51].

## V. Conclusions

The previous three-dimensional color-gradient LB models usually suffer from the lack of Galilean invariance and considerable numerical errors because of the error terms in the recovered macroscopic equations. In this paper we have theoretically analyzed the physical origin of the error terms in the previous models. Based on the theoretical analysis, we have proposed an improved three-dimensional color-gradient LB model for simulating immiscible multiphase flows. Specifically, a high-order term is added to the equilibrium distribution function, through which the off-diagonal elements of the third-order moment of the equilibrium distribution function can satisfy the required relationship for recovering the correct Navier-Stokes equations. Meanwhile, the deviations of the diagonal elements are corrected via a correction term in the LB equation. Compared with the previous models, the present model eliminates the error terms and therefore improves the numerical accuracy and enhances the Galilean invariance.

To validate the proposed color-gradient LB model, numerical simulation have been performed. The test of a moving droplet in a uniform flow field has been employed to verify the Galilean invariance of the improved model. It has been shown that the shape of the droplet becomes elliptic when the original model is used, while the improved model allows the droplet to retain its circular shape. Numerical simulations have also been carried out for the layered two-phase flow and three-dimensional Rayleigh-Taylor instability, which show that the numerical accuracy of the improved model has been significantly improved in comparison with the original model. Furthermore, the capability of the improved color-gradient LB model for simulating dynamic multiphase flows at a relatively large density ratio has been demonstrated by simulating droplet impact on a solid surface.

## Acknowledgments

This work was supported by the National Natural Science Foundation of China (No. 51822606) and the UK Consortium on Mesoscale Engineering Sciences (UKCOMES) under the UK Engineering and Physical Sciences Research Council Grant No. EP/R029598/1.

### Appendix: The transformation matrix $\mathbf{M}$ and the equilibria $\mathbf{m}^{k,eq}$ in Eq. (41)

$$
\mathbf{M}=\begin{bmatrix}
1&1&1&1&1&1&1&1&1&1&1&1&1&1&1&1&1&1&1\\
0&1&-1&0&0&0&0&1&-1&1&-1&1&-1&1&-1&0&0&0&0\\
0&0&0&1&-1&0&0&1&-1&-1&1&0&0&0&0&1&-1&1&-1\\
0&0&0&0&0&1&-1&0&0&0&0&1&-1&-1&1&1&-1&-1&1\\
0&1&1&1&1&1&1&2&2&2&2&2&2&2&2&2&2&2&2\\
0&2&2&-1&-1&-1&-1&1&1&1&1&1&1&1&1&-2&-2&-2&-2\\
0&0&0&1&1&-1&-1&1&1&1&1&-1&-1&-1&-1&0&0&0&0\\
0&0&0&0&0&0&0&1&1&-1&-1&0&0&0&0&0&0&0&0\\
0&0&0&0&0&0&0&0&0&0&0&1&1&-1&-1&0&0&0&0\\
0&0&0&0&0&0&0&0&0&0&0&0&0&0&0&1&1&-1&-1\\
0&0&0&0&0&0&0&1&-1&-1&1&0&0&0&0&0&0&0&0\\
0&0&0&0&0&0&0&1&-1&1&-1&0&0&0&0&0&0&0&0\\
0&0&0&0&0&0&0&0&0&0&0&1&-1&-1&1&0&0&0&0\\
0&0&0&0&0&0&0&0&0&0&0&1&-1&1&-1&0&0&0&0\\
0&0&0&0&0&0&0&0&0&0&0&0&0&0&0&1&-1&-1&1\\
0&0&0&0&0&0&0&0&0&0&0&0&0&0&0&1&-1&1&-1\\
0&0&0&0&0&0&0&1&1&1&1&0&0&0&0&0&0&0&0\\
0&0&0&0&0&0&0&0&0&0&0&1&1&1&1&0&0&0&0\\
0&0&0&0&0&0&0&0&0&0&0&0&0&0&0&1&1&1&1
\end{bmatrix}.
$$

$$
\mathbf{m}^{k,eq}=\mathbf{M}\mathbf{f}^{k,eq}=\begin{bmatrix}
\rho_k\\
\rho_k u_x\\
\rho_k u_y\\
\rho_k u_z\\
3p_k+\rho_k|\mathbf{u}|^2\\
\rho_k\left(2u_x^2-u_y^2-u_z^2\right)\\
\rho_k\left(u_y^2-u_z^2\right)\\
\rho_k u_x u_y\\
\rho_k u_x u_z\\
\rho_k u_y u_z\\
p_k u_y\\
p_k u_x\\
p_k u_z\\
p_k u_x\\
p_k u_z\\
p_k u_y\\
\left[\rho_k(1-\alpha_k-|\mathbf{u}|^2)+2\rho_k c^2(u_x^2+u_y^2)\right]/6\\
\left[\rho_k(1-\alpha_k-|\mathbf{u}|^2)+2\rho_k c^2(u_x^2+u_z^2)\right]/6\\
\left[\rho_k(1-\alpha_k-|\mathbf{u}|^2)+2\rho_k c^2(u_y^2+u_z^2)\right]/6
\end{bmatrix},
$$

where $p_k=0.5c^2\rho_k\left(1-\alpha_k\right)$.